\documentclass[5p,twocolumn]{elsarticle}
\usepackage{fullpage}
\usepackage{amsfonts}
\usepackage{amssymb}
\usepackage{amsmath}
\usepackage{hyperref}
\usepackage{graphicx}
\usepackage{epstopdf}
\usepackage{subfig}
\usepackage[normalem]{ulem}
\begin{document}
\title{Competition between BCS-pairing and ``moth-eaten effect'' \\in BEC-BCS crossover}
\author[uiuc]{Guojun Zhu}
\ead{gzhu1@illinois.edu}
\author[uiuc,upmc]{Monique Combescot}
\ead{Monique.Combescot@insp.jussieu.fr}
\address[uiuc]{Department of Physics, University of Illinois at Urbana-Champaign, 1110 W Green St, Urbana, IL, 61801}

\address[upmc]{Institut des NanoSciences de Paris, Universit\'{e} Pierre et Marie Curie, CNRS, Tour 22, 4 place Jussieu, 75005 Paris }
\newcommand{\vk}{\ensuremath{\mathbf{k}}}
\providecommand{\vr}{\ensuremath{\mathbf{r}}}
\newcommand{\vp}{\ensuremath{\mathbf{p}}}

\providecommand{\comm}[1]{\textit{\scriptsize \uwave{(#1)}}} 
\newcommand{\td}{{\ensuremath{{\text{(2D)}}}}}
\newcommand{\sd}{{\ensuremath{{\text{(3D)}}}}}
\newcommand{\Arctg}{\ensuremath{\text{Arctg}}}

\numberwithin{equation}{section}
\begin{abstract}
We study the change in condensation energy from a single pair of fermionic atoms to a large number of pairs interacting via the reduced BCS potential. We find that the energy-saving due to correlations decreases when the pair number increases because the number of empty states available for pairing gets smaller (``moth-eaten effect"). However, this decrease  dominates the 3D kinetic energy increase of the same amount of noninteracting atoms only when the   pair number is a sizeable fraction of the number of states available for pairing. As a result, in  BEC-BCS crossover of 3D systems, the condensation energy per pair first increases and then decreases with pair number while in 2D, it always is controlled by the ``moth-eaten effect'' and thus simply  decreases.  
\end{abstract}

\maketitle
\section{Introduction}
It was known for a long time that a 3D system with a weak attractive potential cannot sustain a bound state.  Cooper however showed that in the presence of a frozen Fermi core, a pair of electrons with opposite spins can form a bound state with zero total momentum,  no matter how weak the attraction is\cite{Cooper}.  Note that this ``single pair'' state already is a many-body state because, even if the frozen core electrons do not interact, they still, because of  Pauli blocking, provide a finite density of states which is of importance for pairing. Turning to more than one pair is difficult due to the Pauli exclusion principle between a fixed number of paired electrons. To overcome this difficulty, Bardeen, Cooper and Schrieffer proposed an ansatz for the many-body state in the grand canonical ensemble - with  pair number not fixed - which in the presence of a frozen core, leads to an energy lower than the free electron energy, even in the limit of an arbitrarily small potential\cite{BCS}. Gor'kov and Melik-Barkhudarov then showed that the frozen core is not
  mandatory, provided that one uses a renormalized attraction measured through the low energy scattering amplitude\cite{Gorkov}.   Later on, Eagles\cite{Eagle}, Leggett\cite{LeggettCrossover} and also Nozi\`{e}res Schmitt-Rink\cite{Nozieres} extended the BCS  idea to bridge molecular BEC with Cooper pairing. To do it, they vary the potential amplitude while using a BCS-like grand canonical wave function without frozen core, i.e., all $\vk$ states are involved.  Their work raises the complementary question: how, when the potential is fixed but too weak to hold a bound state, the solution evolves from a single unbound pair to a very large number of pairs which always have a bound state solution. The grand canonical nature of the BCS ansatz makes it inherently many-body, so a direct connection to the one-pair solution is not really possible.  

Five years after the BCS milestone paper, Richardson\cite{Richardson1} and Gaudin\cite{gaudin}, succeeded to write the exact eigenstate for $N$  pairs interacting via the reduced BCS potential, in terms of $N$ parameters solution of $N$ coupled equations. The condensation energy obtained from the BCS ansatz, has been recovered in the infinite $N$ limit\cite{Richardson2,Richardson3,Richardson1968}. 

A decade ago, one of us has developed a new framework\cite{CobosonPhysicsReports} for many-body effects between composite bosons. Most of its applications dealt with semiconductor excitons.   Recently, we have extended this framework to Cooper pairs and rederived Richardson-Gaudin equations\cite{CobosonBcsRich}. We have also succeeded to obtain an analytical solution of these equations \cite{moth,combescotBCS} which exactly matches the energy obtained through the BCS ansatz.  The Richardson-Gaudin approach is all the most suitable to investigate the change in condensation energy from 1 to $N$ pairs when the pairing potential stays constant, because it allows us to  handle a fixed number of pairs with Pauli blocking treated exactly.

When one electron pair is added to a system already having $N$ pairs, the Pauli exclusion principle shows up in two different ways.  (i) Pairing (binding) has to use a smaller phase space, so that the energy saving per pair due to the attracting potential must be smaller in the case of $(N+1)$ pairs than for $N$ pairs since they have less freedom to construct the most favorable correlated state.   This binding decrease is the so-called ``moth-eaten effect" \cite{moth}. (ii) In the absence of attraction, the additional fermion pair fills the next $\vk$-level due to Pauli blocking; so its energy also depends on the $N$ other pairs.  Of course, these two effects must be handled self-consistently. However, it is enlightening to separate them in order to build some intuition. The binding energy decrease resulting from the ``moth-eaten effect" driven by Pauli blocking, scales as $N/N_\Omega$, where $N$ is the number of pairs at hands and $N_\Omega$ the maximum number of pairs in the potential layer which scales as the sample volume. By contrast, the average kinetic energy for free pairs scales as $\epsilon_F$, i.e., $(N/L^3)^{2/3}$, where $L^{3}$ is the sample volume in 3D. 
When $N$ is small, this dominates the ``moth-eaten effect"; so, the condensation energy per pair, which results from the energy difference without and with potential, must increases.  For larger $N$ however, the ``moth-eaten effect" dominates and the condensation energy per pair finally decreases.

This understanding points out an important aspect of the BEC-BCS crossover:  It is usually introduced at the two-body level, a bound-state appearing when the attraction passes some threshold. The many-body solution is then seen as the effective potential threshold turning to zero for the system to condense at vanishing potential. The present     work proposes a somewhat different understanding which better bridges $2$-body to $N$-body systems in a BEC-BCS crossover: as the pair number increases, a correlated state develops at a lower but still finite potential.

The paper is organized as follow:

In section \ref{sec:model}, we describe the model. We recall the single pair case and then turn to a qualitative understanding of the many-pair system through the condensation energy change when the pair number increases. We show that this change is quite different in 2D, with a constant density of state, and 3D  with a density of state which cancels at zero energy.  To support this understanding, in section \ref{sec:twoPair}, we carefully study two pairs through the resolution of the corresponding  Richardson-Gaudin equations with a $\sqrt{\epsilon}$ density of state: we show that  a binding indeed develops when turning from one to two pairs for a potential set exactly equal to the threshold value for one pair. We then conclude.

\section{Physical understanding}
\subsection{The model\label{sec:model}}
We consider $N$ pairs of fermionic atoms with creation operators $a_\vk^\dagger$ and $b_\vk^\dagger$, ruled by the hamiltonian
$H=H_{0}+V_{BCS}$. For same mass atoms, the kinetic part $H_0$ reads 
\begin{equation}
H_0=\sum_{\vk}\epsilon_\vk(a^\dagger_\vk{}a^{}_\vk+b^\dagger_\vk{}b^{}_\vk)
\end{equation}
We take as potential  the reduced BCS potential of standard superconductivity, but without its frozen core, namely
\begin{equation}
V_{BCS}=-v\sum_{\vk\vk'}w_{\vk'}w_\vk\beta^\dagger_{\vk'}{}\beta^{}_\vk
\label{eq:VBcs}
\end{equation}
 where $\beta^\dagger_{\vk}=a^\dagger_{\vk}b^\dagger_{-\vk}$ while $w_\vk=1$ for $0<\epsilon_\vk<\Omega$ and zero otherwise; so attraction between zero-moment pairs acts from zero to a sharp cutoff $\Omega$. While this cut-off  bears no connection with phonon energies, we can still  relate it to a physical quantity,  the scattering length, as shown below.
 \subsection{One pair\label{sec:onePair}}
The energy $E_1$ of a single pair in this potential follows from Cooper's equation
\begin{equation}
\frac{1}{v}=\sum_{\vk}\frac{\omega_\vk}{2\epsilon_\vk-E_1}\equiv{}S(E_1)
\label{eq:onePair}
\end{equation}

(i) In 2D, the density of state is constant.  By transforming the sum over $\vk$ into an integral, we get for negative $E$, 
\begin{equation}
S^{(\text{2D})}(E<0)=\rho\int_0^{\Omega}\frac{d\epsilon}{2\epsilon-E}=\frac{\rho}{2}\ln\left(\frac{2\Omega-E}{-E}\right)
\label{eq:s1pair}
\end{equation}
This function tends to infinity when $E\rightarrow{}0_{-}$ and to zero as $\rho\Omega/(-E)$ when $E\rightarrow-\infty$. A bound state, solution of Eq. (\ref{eq:onePair}), thus exists no matter how weak $v$ is. It reads
$
E_1^{(\text{2D})}=-2\Omega\sigma/(1-\sigma)
$
with $\sigma=e^{-2/\rho{v}}$. Note that while $\rho$ increases linearly with sample volume, $\rho{v}$ stays constant.

(ii) In 3D, the density of states can be written as  
$\rho(\epsilon)=\rho\sqrt{\epsilon/\Omega}$
where $\rho$ now is the density of state at the potential upper boundary. So
\begin{equation}
\begin{split}
S^\sd(E<0)&=\rho\int_0^{\Omega}{}d\epsilon\frac{\sqrt{\epsilon/\Omega}}{2\epsilon-E}\\
	&=\rho\left[1-\sqrt{\frac{-E}{2\Omega}}\Arctg\sqrt{\frac{2\Omega}{-E}}\right]
\end{split}
\end{equation}
tends to $\rho$ when $E\rightarrow0_-$ and to zero as $\frac{2}{3}\rho\Omega/(-E)$ when $E\rightarrow-\infty$. 
A bound state thus exists for $v$ larger than a threshold $v_{\text{th}}=1/\rho$.  For a potential just above threshold, the single pair energy tends to zero as 
$
E_1^\sd\approx-8(\rho v-1)^2\Omega/\pi^2
$
while far above threshold
$E_1^\sd\approx-\frac{2}{3}\rho{v}\Omega$

(iii) Using this result, we can relate the s-wave scattering length $a_{s}$, commonly used for cold gases,  to the potential cut-off $\Omega$ via the density of state  at this cut-off, $\rho=mL^3\sqrt{2m\Omega}/2\pi^2$. Indeed, for fermion pairs interacting via $V_{BCS}$, the T-matrix for S-wave  reduces to
\begin{equation}
T^{0}_{k}=\frac{-v\omega_k}{1-vS(2\epsilon_k+i0_+)}
\end{equation}
with, from Eq.(2.5), $S^\sd(2\epsilon_k+i0_+)\simeq\rho(1+i\pi\sqrt{\epsilon_k/4\Omega})$  for $\epsilon_k\ll\Omega$. The scattering length then follows from the scattering amplitude $f^0_k= -a_s/(1+ika_s)$ which depends on the T-matrix as $f^0_k= -mL^3T^{0}_{k}/4\pi$. So, $a_s\simeq mL^3v/4\pi(\rho v-1)$. For $v$ slightly above the single pair threshold $1/\rho $, we find $a_s$ positive with a pair binding energy $E_{b}\approx-1/ma_s^{2}$, while below threshold, $a_s$ is negative and no bound state exists. 

\subsection{N pairs\label{sec:NPair}}
Richardson \cite{Richardson1} and Gaudin \cite{gaudin} showed that the energy of $N$ fermion pairs interacting via $V_ {BCS}$ reads as $E_N=R_1+\cdots+R_N$ where the $R_i$'s follow from $N$ coupled equations
\begin{equation}
\frac{1}{v}=\sum_\vk\frac{w_\vk}{2\epsilon_\vk-R_i}+\sum_{j\neq{}i}\frac{2}{R_i-R_j}
\end{equation}

(i) {\it 2D systems}: Very recently\cite{moth,CombTren}, we have derived a compact solution of these equations when the density of state is constant above a 3D frozen core, as for the $N$ Cooper pairs  in standard BCS superconductivity. Using this result  for 2D systems which have a constant density of states whatever the electron energy is, we get
\begin{equation}\label{eq:E2dN}
 E^\td_N=N\,E^\td_1+\frac{N(N-1)}{\rho}\frac{1+\sigma}{1-\sigma}
\end{equation}
within under-extensive terms in $(N/\rho)^{n}$.
The energy difference without and with potential
leads to a condensation energy per pair $\epsilon_N= \left[E_N(v=0)-E_N\right]/N$ equals  
  \begin{equation}
\epsilon^{(2D)}_N=\left[1-\frac{N-1}{N_\Omega}\right]\frac{2\sigma}{1-\sigma}\Omega\label{eq:E2D}
\end{equation}
 the total number of states in the potential layer being $N_\Omega=\sum_k{}w_{\vk}=\rho\Omega$ in 2D. This shows that $\epsilon^\td_N$  decreases linearly with $N$, due to the ``moth-eaten effect" induced by Pauli blocking on the number empty states feeling the potential and thus available to form the correlated $N$-pair state.  Note that for complete filling, $
\epsilon^\td_{N_\Omega}=[\frac{2\sigma}{1-\sigma}]/\rho$ goes to zero as $1/\rho$: for $N=N_\Omega$, the system has lost all its freedom to construct a lower energy state.

(ii) {\it 3D systems}: A compact expression of the $N$-pair energy is not known for a $\sqrt{\epsilon}$ density of states. We can yet say that, due to the same decrease of available states for pairing, the ``moth-eaten effect" must bring the condensation energy per pair down to zero when $N$ approaches the total number of pairs in the potential layer, which in 3D, reads  $N_\Omega=\sum_{\vk}w_{\vk}=2\rho \Omega/3$.

\begin{figure}[htb]
	\centering
		\includegraphics[width=0.8\columnwidth]{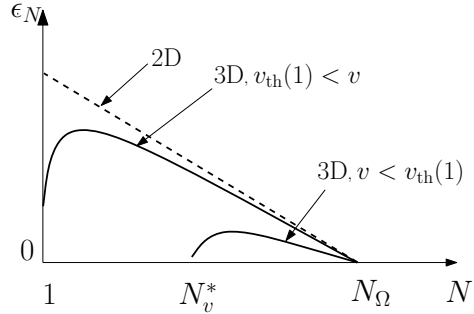}
	\caption{Condensation energy per pair as a function of pair number: in 2D (dashed line), it always decreases due to the ``moth-eaten effect" while in 3D (solid lines for  $v$ below or above threshold for binding a single pair $v_{\text{th}}(1)=1/\rho$), this effect dominates for large $N$. As a result, in BEC-BCS crossover the condensation energy per pair has a maximum which results from the competition between moth-eaten effect and kinetic energy increase when adding a pair.}
	\label{fig:3dCondChange}
\end{figure}

$\bullet$ For $v$ smaller than $v_{th}(1)=1/\rho$, the condensation energy per pair $\epsilon_N^\sd$ is equal to zero for $N=1$ and also for $N=N_\Omega$ due to the ``moth-eaten effect".  It is however clear that $\epsilon_N^\sd$ cannot stay equal to zero for all $N$ because when $N$ gets large, we can always think of freezing $N_0$ of the $N$ electrons as in standard BCS superconductivity. The density of states for the other $(N_\Omega - N_0)$ states in the potential layer is then finite; So, the $(N-N_0)$ pairs in this layer can always condense, no matter how weak $v$ is. For $N_{0}\gg{}N_{\Omega}-N_{0}$, (a thin shell over a large core in momentum space), the density of state in the shell can be taken constant as in 2D. We can then use Eq.(\ref{eq:E2D}) to estimate the total binding energy of these $(N-N_0)$ pairs which only enjoy $(N_\Omega - N_0)$, instead of $N_{\Omega}$  states, for pairing. This gives
\begin{equation}\label{eq:E3D}
{{E}}_N^{(3D)}(N_0)=(N-N_0)\left[1-\frac{(N-N_0)-1}{N_\Omega-N_0}\right]\frac{2\bar\sigma}{1-\bar\sigma}\Omega
\end{equation}
with $\bar{\sigma}=e^{-2/{\bar{\rho}v}}$ where $\bar\rho$ is the average density of states above the frozen core, $\rho(N_0/N_\Omega)^{ 1/3}<\bar\rho<\rho$. By taking for  $\bar\rho$ its lower boundary, we find that ${{E}}_N^{(3D)}(N_0)$ has a maximum which results from a competition when $N_0$ increases, between the ``moth-eaten effect" and the increase of the energy contribution $2\bar\sigma\Omega/(1-\bar\sigma)$  from each available states for pairing.
This argument shows that for $v$ smaller than the threshold for one pair, but $N$ large enough, a BCS-like collective state must develop with a non-zero condensation energy. 
As a result, the threshold potential  $v_{th}(N)$ for the appearance of $N$-pair condensation  must decrease with $N$, down to essentially zero when $N$ reaches a sizable fraction of the total number of pairs $N_\Omega$ in the potential layer.  Consequently, the condensation energy per pair $\epsilon^{\sd}_{N}$ stays equal to zero up to $N=N_v^*$ reached for $v=v_{th}(N_v^*)$. It then increases and ultimately decreases to zero for $N=N_\Omega$ (see Fig. \ref{fig:3dCondChange}), due to the moth-eaten effect which always dominates when most of the pair states available for pairing are occupied.  (Since the $N$ dependence of $\epsilon_{N}^{\sd}$ above $N_{v}^{*}$ is not yet known, we have preferred not to show it on Fig \ref{fig:3dCondChange}.)

$\bullet$  For $v$ above $v_{th}(1)$, a finite condensation energy already exists for $N=1$. Continuity with $v<v_{th}(1)$ leads us to think that $\epsilon^\sd_N$ must first increase with $N$ and then decrease due to the same moth-eaten effect which dominates condensation when $N$ approaches full-filling. 
To physically understand this behavior, we can note that the condensation energy results from a difference between the free pair and the correlated pair energy, which both increase with $N$ due to Pauli blocking.  The question then is to understand why, for small $N$, the free pair energy increases faster than the correlated pair energy. When one free pair is added at the energy level $\epsilon$, the kinetic energy increases by $2/\rho(\epsilon)$ since the fermion density of state is by definition related to the energy increase through $\rho(\epsilon)=\Delta{}N/\Delta\epsilon$.  In 3D, with a $\sqrt{\epsilon}$ density of states, the energy change when going from $N$ to $N+1$ pairs, is thus larger at low energy. By contrast, correlated pairs are made with all pair states between $0$ and $\Omega$. When the pair number increases from $N$ to $N+1$, a very small fraction of each of these ($0,\Omega$) states is blocked, so that the kinetic energy change is far smaller than when blocking a single low energy state.  Consequently, when $N$ is small,  the energy difference between adding one free pair and one correlated pair must increase with $N$. By contrast, when $N$ gets large, the energy cost $1/\rho(\epsilon)$ to add one free pair is essentially equal to the cost to block most $\vk$ states making the correlated pairs.  We are then left with the ``moth-eaten effect" on the bound state itself and the condensation energy decreases. 

We wish to stress that this understanding is fully supported by the monotonous decrease  found for $\epsilon^\td_N$: indeed, the 2D density of state being constant, the energy cost to block a low energy state is the same as the one to block  any other states making the correlated pairs. The ``moth-eaten effect" controls  the whole  behavior and the condensation energy per pair always decreases when $N$ increases.

We here use a sharp potential cutoff $\Omega$ for convenience.  Real systems do not have such a cutoff; so,   the fact that the condensation energy vanishes when all states below $\Omega$ are filled, can be questioned.  A sharp cutoff $\Omega$ mimics the fall-off of the potential at high energy.  The potential amplitude defines an energy scale, which in turn defines a phase space.  Beyond this space, the potential is too weak to substantially affect the fermion distribution which just behaves as free, with a zero condensation energy. 

To establish this qualitative understanding on stronger  grounds, let us concentrate on  $2$ pairs. The change from 1 to 2 pairs will give us the trend for the $N$-dependence of $\epsilon^{(3D)}_N$ since, for $N=2$, Pauli blocking which drives this $N$-dependence, is already present.  

\section{Condensation energy for two pairs\label{sec:twoPair}}
According to Richardson-Gaudin, the exact eigen-energy of two Cooper pairs reads as $E_2=R_1+R_2$ with $(R_1,R_2)$ solution of
\begin{equation}
\frac{1}{v}=\sum_{\vk}\frac{w_\vk}{2\epsilon_\vk-R_1}+\frac{2}{R_1-R_2}=(R_{1}\leftrightarrow{}R_{2})
\label{eq:richardsonEq}
\end{equation}

\subsection{2D systems}
We have solved these coupled equations analytically for a constant density of states above a frozen core\cite{combescotBCS}.  Using  this work  for 2D systems with a constant density of states $\rho$  between $0$ and $\Omega$, we get the energy difference between two correlated pairs and two single pairs in the large sample limit as 
\begin{equation}
E^{\td}_2-2E_1^{\td}=\frac{2}{\rho}\left(1+\frac{2\sigma}{1-\sigma}\right)+\text{O}(\frac{1}{\rho^2})
\end{equation}
 Since $2/\rho$  is the exact kinetic energy cost to go from 1 to 2 pairs when the density of state is constant, we find a binding energy decrease when going from one to two pairs equal to $2\sigma/\rho(1-\sigma)$ in agreement with Eq. \ref{eq:E2dN}, taken for $N=2$, as a consequence of the ``moth-eaten effect".

\subsection{3D systems}

The situation is more complex in 3D because the potential has to be higher than a threshold to sustain a bound state for one pair.

(i) {\it Potential above threshold} $v>v_{th}(1)=1/\rho$. A single pair  has a bound state  with $E_1$  finite negative. The energy change between two correlated pairs $R_{1}+R_{2}$ and two single pairs  $2E_{1}$ can only come from Pauli blocking due to the very peculiar form of the reduced BCS potential \cite{moth}. We thus expect $R_1+R_2\approx2E_1$ for $L^3\rightarrow\infty$, i.e., $\rho\rightarrow\infty$. This leads us to expand the sum of Eq.(\ref{eq:richardsonEq}) in terms of $(R_{1}-E_{1})$.  Using Eq.(\ref{eq:onePair}), we get
\begin{equation}
0=\sum_{n=1}^{\infty}(R_{1}-E_{1})^{n}\sum_{\vk}\frac{w_\vk}{(2\epsilon_\vk-E_1)^{n+1}}+\frac{2}{R_1-R_2}
\end{equation}
The sum over $\vk$, calculated with a $\rho\sqrt{\epsilon/\Omega}$ density of state, gives
\begin{equation}
\sum_{\vk}\frac{w_\vk}{(2\epsilon_\vk-E_1)^{n+1}}=\\\frac{\rho}{(-E_{1})^{n}}\sqrt{\frac{-E_{1}}{2\Omega}}K_{n}(\frac{2\Omega}{-E_{1}})
\end{equation}
where $K_{n}(x)$, defined as
\begin{equation}
K_{n}(x)\equiv\int_{0}^{x}\frac{\sqrt{y}\;dy}{2(y+1)^{n+1}}
\end{equation}
goes to a finite value $K_{n}$ when $x$ goes to infinity, which is the relevant limit since $|E_1|\ll\Omega$.
By setting
$R_{i}=E_{1}(1-t_{i})$, Richardson-Gaudin equations (\ref{eq:richardsonEq})  then lead to
\begin{equation}
0=\sum_{n=1}^{\infty}t_{1}^{n}K_{n}+\frac{1}{\rho\Omega}\left(\frac{2\Omega}{-E_{1}}\right)^{3/2}\frac{1}{t_1-t_2}=(t_{1}\leftrightarrow{}t_{2})
\end{equation}
The sum and difference of these two equations give
\begin{gather}
0=\sum_{n=1}^{\infty}(t_{1}^{n}+t_{2}^{n})K_{n}\label{eq:t2}\\
-\lambda^{2}=(t_{1}-t_{2})\sum_{n=1}^{\infty}(t_{1}^{n}-t_{2}^{n})K_{n}\label{eq:t1}
\end{gather}
where $\lambda^2=(\frac{2}{\rho\Omega})(\frac{2\Omega}{-E_{1}})^{3/2}$.
Two different regimes thus appear: 

$\bullet$ For $\lambda^{2}\ll1$, Eq.(\ref{eq:t1}) gives $(t_{1}-t_{2})^{2}\approx-\lambda^{2}/K_{1}$. Since $t_{1}^{2}+t_{2}^{2}$ is just $\left[(t_{1}+t_{2})^{2}+(t_{1}-t_{2})^{2}\right]/2$,  Eq.(\ref{eq:t2}) gives $(t_{1}+t_{2})\approx\lambda^{2}K_{2}/2K_{1}^{2}$.  So the two correlated pair energy  reads for $\rho$ large as 
\begin{equation}
E_{2}=R_{1}+R_{2}\approx2\left(E_{1}+\frac{1}{\rho}\sqrt{\frac{2\Omega}{-E_{1}}}\frac{K_{2}}{K_{1}^{2}}\right)
\end{equation}
$E_{2}$ thus increases with sample volume as $1/\rho\sim1/L^{3}$.
To get the condensation energy change from one to two pairs, we must compare this increase with the one of two free pairs.  Due to Pauli blocking, the second pair momentum must be $2\pi/L$; so, the kinetic energy increase for two free pairs scales as $1/L^{2}$, which for large $L$, is larger than the two correlated pair increase. 
The condensation energy per pair, which is the energy difference without and with potential, thus increases from 1 to 2 pairs - in agreement with the continuity argument of section 2. 

 $\bullet$ The regime $\lambda^{2}$ of the order or larger than $1$ corresponds to $E_{1}$ small, more precisely $-E_{1}<\Omega^{1/3}/\rho^{2/3}\approx1/mL^{2}$.  The single pair binding energy is then smaller than the kinetic energy difference between the two lowest free states: in this limit, the discrete sum over $\vk$ cannot be replaced by an integral and the above procedure fails.  This small $E_1$ regime must be handled along  the $E_1=0$ case studied now.

(ii) {\it Potential at threshold} $v=1/\rho$. The single pair energy $E_1$ then cancels; so we cannot rescale the $R_i$'s in terms of $E_1$ as previously done. It is reasonable to think that, as for $E_1$ small, the threshold behavior for $E_1=0$, cannot be derived by replacing the sum over $\vk$ by an integral.
To prove it, let us first subtract the two Richardson-Gaudin equations (\ref{eq:richardsonEq}). This gives
\begin{equation}
-\frac{4}{(R_1-R_2)^2}=\sum_{\vk}\frac{w_\vk}{(2\epsilon_\vk-R_1)(2\epsilon_\vk-R_2)}\label{eq:2PairMinus}
\end{equation}
For $R_1+R_2=E_2=2R$ and $R_1-R_2=2iR'$ with $R$ real but $R'$ a priori complex, we get
\begin{equation}
\frac{1}{R'^2}=\sum_{\vk}\frac{w_\vk}{(2\epsilon_\vk-R)^2+R'^2}
\end{equation}
which also reads 
\begin{multline}
R'^{2}\left[\frac{1}{|R'|^{4}}-\sum_{\vk}\frac{w_{\vk}}{|(2\epsilon_{\vk}-R)^{2}+R'^{2}|^{2}}\right]\\
=\sum_{\vk}\frac{w_{\vk}(2\epsilon_{\vk}-R)^{2}}{|(2\epsilon_{\vk}-R)^{2}+R'^{2}|^{2}}
\end{multline}
This imposes $R'^2$ real, i.e., $(R_1,R_2)$ both real or complex conjugate.

$\bullet$ Let us first show that $(R_1,R_2)$ cannot be complex conjugate. For that, we use Eq.(\ref{eq:onePair}) for $E_1=0$. Eq.(\ref{eq:richardsonEq}) then gives
\begin{equation}
-\frac{2}{R_{1}(R_{1}-R_{2})}=\sum_\vk\frac{w_{\vk}}{2\epsilon_{\vk}(2\epsilon_{\vk}-R_{1})}
\end{equation}
If $(R_1,R_2)$ were complex conjugate, we could set $R_1=R^*_2=2 \Omega r e^{i\theta}$ with $r$ real and $0\leqslant\theta<\pi$. For $(R_1,R_2)$ far enough from the real axis to possibly replace the $\vk$ sum by an integral, we get
\begin{multline}\label{eq:r1r2}
\frac{1}{2\rho\Omega{r^{3/2}}}\left(\frac{1}{\cos\theta/2}+\frac{1}{\sin\theta/2}\right)\\
=\log\left(\frac{1-\sqrt{r}e^{i\theta/2}}{-\sqrt{r}e^{i\theta/2}}\frac{\sqrt{r}e^{i\theta/2}}{1+\sqrt{r}e^{i\theta/2}}\right)
\end{multline}
We expect a 2-pair energy $E_{2}$ far smaller than the potential threshold,  $|R|\ll2 \Omega$, i.e., $r\ll1$. The R.H.S. of Eq. (\ref{eq:r1r2}) then reduces to 
$(i\pi-2\sqrt r\cos\theta/2)(1+O(\sqrt r))$. The real part of the above equation then gives $-1\approx4 \rho \Omega\cos^2\theta/2$ which is impossible.

$\bullet$ We now show that $(R_1,R_2)$ cannot be both real outside $(0,\Omega)$. For that, we add the two Richardson-Gaudin equations (\ref{eq:richardsonEq}) and use $1/v=\sum_{\vk}w_\vk/2\epsilon_\vk$ at threshold. This gives
\begin{equation}
0=\sum\frac{w_{\vk}}{2\epsilon_{\vk}}\left[\frac{R_{1}}{2\epsilon_{\vk}-R_{1}}+\frac{R_{2}}{2\epsilon_{\vk}-R_{2}}\right]
\end{equation}
If $(R_1,R_2)$ were both real outside $(0,\Omega)$, each term of the bracket would be negative; so, the above equation is not fulfilled.

$\bullet$ We are thus left with $(R_1,R_2)$ both real, with one at least in $(0,\Omega)$ - or possibly $(R_1,R_2)$ complex conjugate but very close to the real axis so that the $\vk$ sum cannot be replaced by an integral. To estimate the two pairs ground state, we can reduce this discrete $\vk$ sum to its first  few terms, namely $\vk=0$, i.e., $\epsilon_\vk=0$ and
 $(\vk_x,\vk_y,\vk_z)=\pm 2\pi/L
 $, i.e., $\epsilon_\vk=(\pm 2\pi/L)^2/2m=\epsilon_L$. The Richardson-Gaudin equations with these seven terms in the sum reads at threshold
\begin{equation}
\begin{split}
\rho=\frac{1}{v}&=\frac{1}{-R_{1}}+\frac{6}{2\epsilon_{L}-R_{1}}+\frac{2}{R_{1}-R_{2}}\\
&=(R_{1}\leftrightarrow{}R_{2})
\end{split}
\end{equation}
The proper rescaling then is  $R_i=2\epsilon_L u_i$. So the two Richardson-Gaudin equations  read
\begin{equation}
\begin{split}
-\frac{1}{u_{1}}+\frac{6}{1-u_{1}}+\frac{2}{u_{1}-u_{2}}&=2\rho\epsilon_{L}\\
-\frac{1}{u_{2}}+\frac{6}{1-u_{2}}+\frac{2}{u_{2}-u_{1}}&=2\rho\epsilon_{L}
\end{split}\label{eq:t12}
\end{equation}
where $\rho\epsilon_L$ scales as $L$.

-- We first look for $(u_1,u_2)$ both real. For $L$ large, we get $-1/u_1\approx 2 \rho\epsilon_L \approx 6/(1-u_2)$. This gives $R_1=-1/\rho$ and $R_2=2 \epsilon_L-6/\rho$; so, the two-pair energy reduces to $R_1+R_2\approx2 \epsilon_L-7/\rho$ .

--  We then look for $(u_1,u_2)$ complex conjugate, i.e., $u_1=u+iu'=u^*_2$. It is possible to show that Eqs.(\ref{eq:t12}) then have no solution for $u<u'$ or $u\simeq u'$ while a solution exists for $u'>u$. It reads $u\approx1-5/2\rho\epsilon_L$ with $u'\propto1/\rho\epsilon_L$. This leads to a 2-pair energy $R_1+R_2\approx4 \epsilon_L-10/\rho$ which is above the energy found for $(R_1,R_2)$ real.

Since the kinetic energy of two free pairs is $2\epsilon_L$, the condensation energy for two pairs at the potential threshold for one pair , $v=1/\rho$, reduces to 
$ 2\epsilon_L-(2 \epsilon_L-7/\rho)=+7/\rho$. Since this condensation energy  is zero for one pair, it indeed increases  from $N=1$ to $N=2$ in agreement with the continuity argument of section 2. As a result, the potential $v_{th}(1)=1/\rho$, too weak to sustain a bound state for a single pair, can bind two pairs: by continuity, the potential threshold for pairing, $v_{th}(N)$, must decrease when the pair number increases.

\section{Conclusion\label{sec:conclusion}}
We  show that a system of fermion pairs with attraction too weak to sustain a bound state, can nevertheless form a many-body correlated state when the pair number increases.  Using Richardson-Gaudin equations, we demonstrate that the required potential strength for condensation decreases as the pair number increases.  For strong attraction, two fermions form a bound state; the many-body state can then be viewed as a congregation of single pairs with Pauli blocking decreasing their binding energy (``moth-eaten effect").  However, when attraction gets weaker, more than one pair is necessary to form a condensed state as explicitly shown by going from one pair to two pairs at the potential threshold for binding a single pair.  For a vanishing interaction, a large number of pairs are required to obtain BCS condensation.   A cross-over then appears when the pair number increases at constant but finite potential.

 This work also reveals, through difference between 2D and 3D behaviors, the subtle interplay which exists between Pauli blocking at the free fermion level which increases the kinetic energy when the pair number increases, and Pauli blocking at the paired level through the ``moth-eaten effect" resulting from the decrease of the number of empty states available for pairing.

M.C. wishes to thank the Institute for Condensed Matter Physics of the University of Illinois at
Urbana-Champaign, and Tony Leggett in particular, for enlightening discussions during her various invitations which led to the present work. G.J.Z. is supported partly by NSF under grant No. DMR 09-06921.  

\bibliographystyle{model1a-num-names}

\end{document}